\documentclass[%
reprint,
superscriptaddress,
amsmath,amssymb,
prb,
]{revtex4-1}

\usepackage{graphicx}
\usepackage{dcolumn}
\usepackage{bm}
\usepackage{hyperref}
\usepackage[mathlines]{lineno}
\usepackage{siunitx}
\usepackage{xspace}
\usepackage{epstopdf}
\usepackage[]{pgfplots}
\pgfplotsset{compat=1.12}
\usetikzlibrary{arrows}

\usepackage[eulergreek]{sansmath}

\DeclareSIUnit\decade{dec}

\usepackage[]{natbib}

\usepackage[caption=false]{subfig}

\newcommand{\YMO}{YMnO$_3$\xspace}
\newcommand{\tc}{$T_\mathrm{c}$\xspace}

\newcommand{\ie}{\textit{i.e.,}\xspace}

\begin{document}
	
	\preprint{APS/123-QED}
	
	\title{Local control of improper ferroelectric domains in \YMO }
	
	\author{Lukas Kuerten}
	\email{lukas.kuerten@mat.ethz.ch}
	\affiliation{Departement of Materials Science, ETH Zurich, 8093 Zurich, Switzerland}%
	\author{Stephan Krohns}
	\affiliation{Experimental Physics V, University of Augsburg, 86135 Augsburg, Germany}%
	\author{Peggy Schoenherr}
	\affiliation{Departement of Materials Science, ETH Zurich, 8093 Zurich, Switzerland}%
	\author{Katharina Holeczek}
	\affiliation{Experimental Physics V, University of Augsburg, 86135 Augsburg, Germany}%
	\author{Ekaterina Pomjakushina}%
	\affiliation{Laboratory for Multiscale Materials Experiments, Paul Scherrer Institut, 5232 Villigen, Switzerland}
	\author{Thomas Lottermoser}
	\author{Morgan Trassin}
	\affiliation{Departement of Materials Science, ETH Zurich, 8093 Zurich, Switzerland}%
	\author{Dennis Meier}%
	\affiliation{Department of Materials Science and Engineering, Norwegian University of Science and Technology NTNU, 7491 Trondheim, Norway}
	\author{Manfred Fiebig}
	\affiliation{Departement of Materials Science, ETH Zurich, 8093 Zurich, Switzerland}%
	
	%
	%
	
	\date{\today}
	\begin{abstract}
		
		Improper ferroelectrics are described by two order parameters: a primary one, driving a transition to long-range distortive, magnetic or otherwise non-electric order, and the electric polarization, which is induced by the primary order parameter as a secondary, complementary effect. Using low-temperature scanning probe microscopy, we show that improper ferroelectric domains in \YMO can be locally switched by electric field poling. However, subsequent temperature changes restore the as-grown domain structure as determined by the primary lattice distortion. The backswitching is explained by uncompensated bound charges occuring at the newly written domain walls due to the lack of mobile screening charges at low temperature. Thus, the polarization of improper ferroelectrics is in many ways subject to the same electrostatics as in their proper counterparts, yet complemented by additional functionalities arising from the primary order parameter. Tailoring the complex interplay between primary order parameter, polarization, and electrostatics is therefore likely to result in novel functionalities specific to improper ferroelectrics. 
		
		
	\end{abstract}

	\maketitle
	
	
	\section{Introduction}

	In improper ferroelectrics, the spontaneous polarization emerges as subordinate effect to a primary order parameter which can be a lattice distortion, a magnetization or another non-electric quantity \cite{Levanyuk1974, Kimura2003, Aken2004, Ikeda2005}. This dependence can lead to properties not observed in their polarization-driven proper ferroelectric counterparts. Improper ferroelectrics can be expected to be more robust towards extrinsic influences, for example depolarizing fields, allowing domain configurations with unusual head-to-head or tail-to-tail polarization geometries at the domain walls \cite{Choi2010, Jungk2010}. Such domain configurations can have technologically highly relevant properties, ranging from  local conductance enhancement\cite{Meier2012, Oh2015,  McQuaid2017} to functionalities of  advanced circuit elements\cite{Mundy2017, Schaab2018}. 
	
	Both order parameters of improper ferroelectrics, the primary one and the induced polarization, can in principle influence the domain structure, but while it appears obvious that the primary order parameter sets the initial domain structure when crossing the transition temperature,  the role played by the secondary order parameter and the associated electrostatics is not as clear. In the case of the hexagonal manganites ($R$MnO$_3$, with $R$ = Sc, Y, In, Dy -- Lu), one of the most established classes of improper ferroelectrics,  a lattice-trimerizing distortion as primary order parameter  dominates the formation of domains, but only the secondary order parameter is susceptible to controlled poling in an electric field\cite{Lilienblum2015}. Therefore,  a key question is if, and how, domains associated with the primary order parameter formed at the transition temperature \tc may differ from those created by electric field poling of the secondary order parameter within the ordered phase far below \tc. Even though a detailed understanding of this complex interrelation is crucial for  the functionalization of  improper ferroelectrics, this aspect has not received much attention.

	%
	%

	\begin{figure*}
		\centering
		\includegraphics[width=\textwidth]{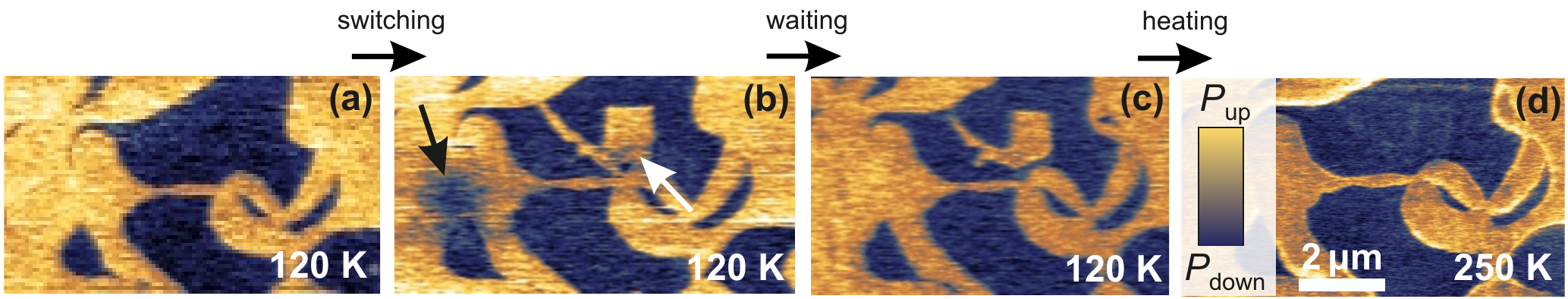}
		\caption{Creation  of improper ferroelectric domains in \YMO by local   low-temperature electric-field poling. a) Pristine domain structure measured by PFM at \SI{120}{\kelvin}. b) A square-shaped area (white arrow) of reversed polarization is created by scanning  while applying +\SI{45}{\volt}  to the AFM tip in contact with the surface. The bright line protruding from the lower end of the square was caused by moving the AFM tip into position for the  poling. When the same voltage is applied to an area polarized in the direction of the applied voltage, surface charging results in a diffuse change of contrast (black arrow). c) The same area of the sample surface imaged after remaining at \SI{120}{\kelvin}  for  6 days after the poling. The artificially created domain is still present, whereas  the  surface-charged area has disappeared. d) At \SI{250}{\kelvin} the domain structure  immediately reverts to the original configuration in (a).}
		\label{fig:square}
	\end{figure*}

	Here, we investigate electric-field poling at the nanometer scale in hexagonal \YMO. 
	In this material, uniform  tilting of the MnO$_5$ bipyramids in the unit cell and a concomitant shift of the ytttrium ions occur at  \SI{1258}{\kelvin}. This lattice-trimerizing distortive transition  drives an improper ferroelectric polarization of \SI{5.6}{\micro \coulomb \per \centi \meter \squared} along the $c$-axis \cite{Fiebig2002,Lottermoser2002,Kimura2003,Fennie2005,Jungk2010}. The resulting domain structure consists of six trimerization-polarization domain states forming vortex-like meeting points with alternating polarization around the vortex core\cite{Choi2010, Jungk2010, Meier2017}.
	
	We use  atomic force microscopy (AFM) to apply local electric fields at cryogenic temperatures, where non-intrinsic effects due to barrier layer capacitances are negligible\cite{Ruff2017, Ruff2018},  creating polarization domains at the nanoscale.  We compare  these written domains to the domains formed via the primary order parameter at \tc. We find that despite the secondary nature of the electric polarization, this polarization dominates the poling behavior just as in conventional ferroelectrics. Domains can be created at will by locally applied electric-fields. However, thermal annealing cycles return the samples to the as-grown domain configuration. This recovery is explained by uncompensated bound charges at the domain walls and the surface, which arise due to decreasing availability  of mobile carriers in the semiconductor at low temperature.
	Hence, despite the secondary nature of the ferroelectric order, the electrostatic conditions overrule the primary lattice trimerization. 
	Quite strikingly, we thus find that improper ferroelectrics retain key characteristics of proper ferroelectrics, yet complemented by functionalities introduced by the secondary nature of the electric order.

	\section{Methods}

	Experiments were performed on  \YMO single crystals grown by the floating-zone method \cite{Meier2017, Roessli2005}. The crystals were cut into platelets with a  thickness of approximately \SI{500}{\micro \meter}  perpendicular to the crystallographic $c$-axis. They were  lapped with an Al$_2$O$_3$ solution and polished with silica slurry, yielding a surface roughness of approximately \SI{1}{\nano \meter}, suitable for AFM measurements. We thus obtained out-of-plane-polarized samples whose trimerization-polarization domains at the surface are separated by nominally uncharged \ang{180} side-by-side domain walls\cite{Choi2010, Jungk2010, Meier2012}. 
	
	Dielectric measurements were performed using a Novocontrol Alpha analyser (at \SI{1}{\hertz} to \SI{1}{\mega \hertz}) and a TF2000 Aixacct system (hystersis loops, at \SI{1}{\hertz}) in combination with a high-voltage booster for voltages up to \SI{2}{\kilo \volt}. Measurements were conducted at \SIrange{50}{300}{\kelvin} in a closed-cycle refrigerator with samples in vacuum to avoid electrical discharge. The properties of semiconducting materials are often superimposed by extrinsic barrier layer contributions\cite{Lunkenheimer2009} which may affect polarization measurements\cite{Scott2007, Loidl2008}.   For  \YMO, a temperature of \SI{120}{\kelvin}  and a frequency of \SI{1}{\hertz} avoid barrier layer capacitances and allow detecting the genuine ferroelectric properties of the material both in bulk and AFM experiments\cite{Ruff2017, Ruff2018}. 
	
	AFM measurements  were performed at \SIrange{120}{250}{\kelvin} in an attoLiquid 2000  AFM setup (attocube GmbH, Germany) with ANSCM-PT Pt/Ir-coated conductive tips (AppNano Inc., USA) in two different modes: 
	
On the one hand, we directly imaged the distribution of the polarization  by piezoresponse force microscopy (PFM)\cite{Kalinin2007, Soergel2011,Gruverman2019}. In this mode, the AFM tip is brought into contact with the sample surface and an AC voltage is applied to the tip. The AFM detects the contraction and expansion of the sample due to the piezoelectric effect. A contraction in phase or  in antiphase with the excitation voltage corresponds to polarization in the upward or downward direction, respectively. 
	
On the other hand, we used the presence of uncompensated charges on the surface to image the domain structure  by  electrostatic force microscopy (EFM) \cite{Kalinin2001, Schoenherr2019}. Due to a difference in strength of  the pyroelectric effect,   the  surface charge differs between domains and domain walls, which is detected as contrast in EFM measurements. For details of the EFM measurement procedure and temperature sequences, see Supplementary Information.
	
Local domain switching was achieved at \SI{120}{\kelvin} by applying DC bias voltages to the AFM tip in contact with the sample surface.

	
	\section{Results}
	
		\begin{figure*}	
		\centering
		\includegraphics[width=0.75\textwidth]{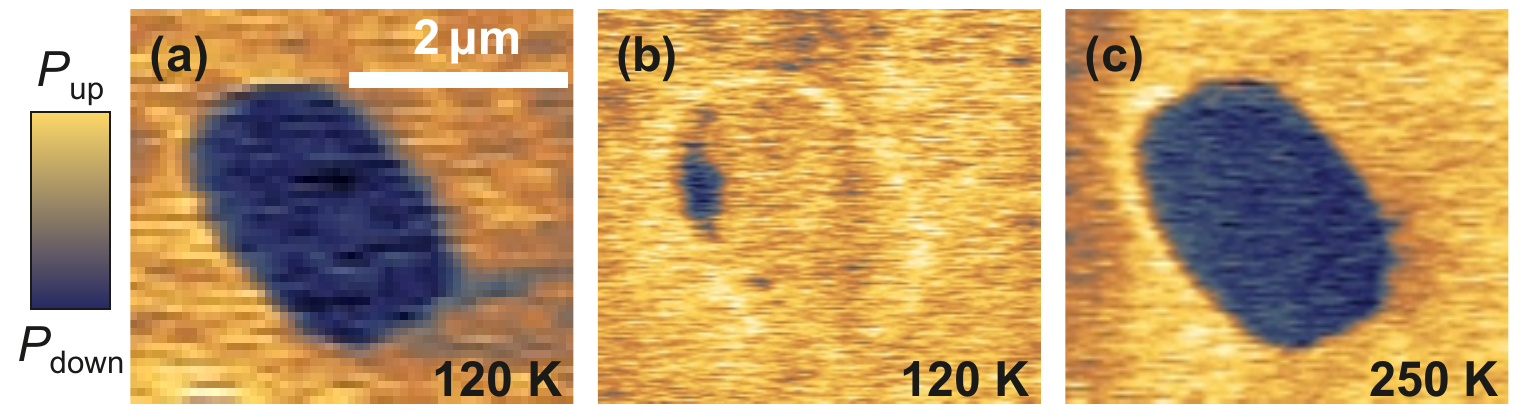}
		\caption{Elimination of improper ferroelectric domains in \YMO by local  low-temperature electric-field poling. a) A down-polarized bubble domain in an up-polarized environment measured by PFM at \SI{120}{\kelvin}. b) The polarization of the bubble domain is reversed by scanning a window of \SI{2}{\micro \meter}$\times$\SI{2}{\micro \meter} covering the bubble with  +\SI{45}{\volt} applied to the AFM tip. Note that the  outline of the original domain is still  visible in the PFM image. c) When increasing the temperature to \SI{250}{\kelvin}, the original bubble domain in (a) is reestablished.}
		\label{fig:bubble}
	\end{figure*}

	PFM measurements at \SI{120}{\kelvin}  showed the typical   trimerization-induced improper ferroelectric domain structure of the hexagonal manganites (Fig.~\ref{fig:square}\,a)). We then created a new domain by applying a voltage of +\SI{45}{\volt} to the AFM tip while scanning a window of \SI{1}{\micro \meter}$\times$\SI{1}{\micro \meter} (Fig.~\ref{fig:square}\,b)). This resulted  in a  square-shaped domain of upwards polarization within a down-polarized domain (white arrow).   The black arrow points to a region where the same poling procedure was applied to an area which was already polarized upwards. The latter led to the injection of surface charges, visible as a diffuse dark region. At \SI{120}{\kelvin}, the written domain  was stable  over a period of more than six days, whereas the space charges disappeared within a few hours (Fig.~\ref{fig:square}\,c)). Finally, we found that when the sample was heated to \SI{250}{\kelvin}, the domain structure reverted to its original configuration, \ie the electric-field-induced square domain disappeared (Fig~\ref{fig:square}\,d)).

	In order to investigate how the ferroelectric domain structure reverts to its previous configuration, we recorded a series of PFM images at higher spatial resolution. Figure \ref{fig:bubble}\,a) shows a down-polarized bubble domain within an up-polarized environment. After  scanning a window of \SI{2}{\micro \meter}$\times$\SI{2}{\micro \meter} covering the entire bubble  with  +\SI{45}{\volt} applied to the tip, the  polarization was mostly reversed so that the bubble disappeared. A faint outline, however, was still observable where the previous as-grown domain wall had been located (Fig.~\ref{fig:bubble}\,b)). This outline was possibly caused by the presence of oxygen interstitials, which are known to accumulate at neutral walls\cite{Schaab2018}, but are immobile at low temperature\cite{Remsen2011} and hence cannot follow the displacement of the domain wall.   When the sample was heated to \SI{250}{\kelvin}, the original domain structure was recovered as depicted in Fig.~\ref{fig:bubble}\,c).

\begin{figure}
		\centering
		\includegraphics[width=0.8\columnwidth]{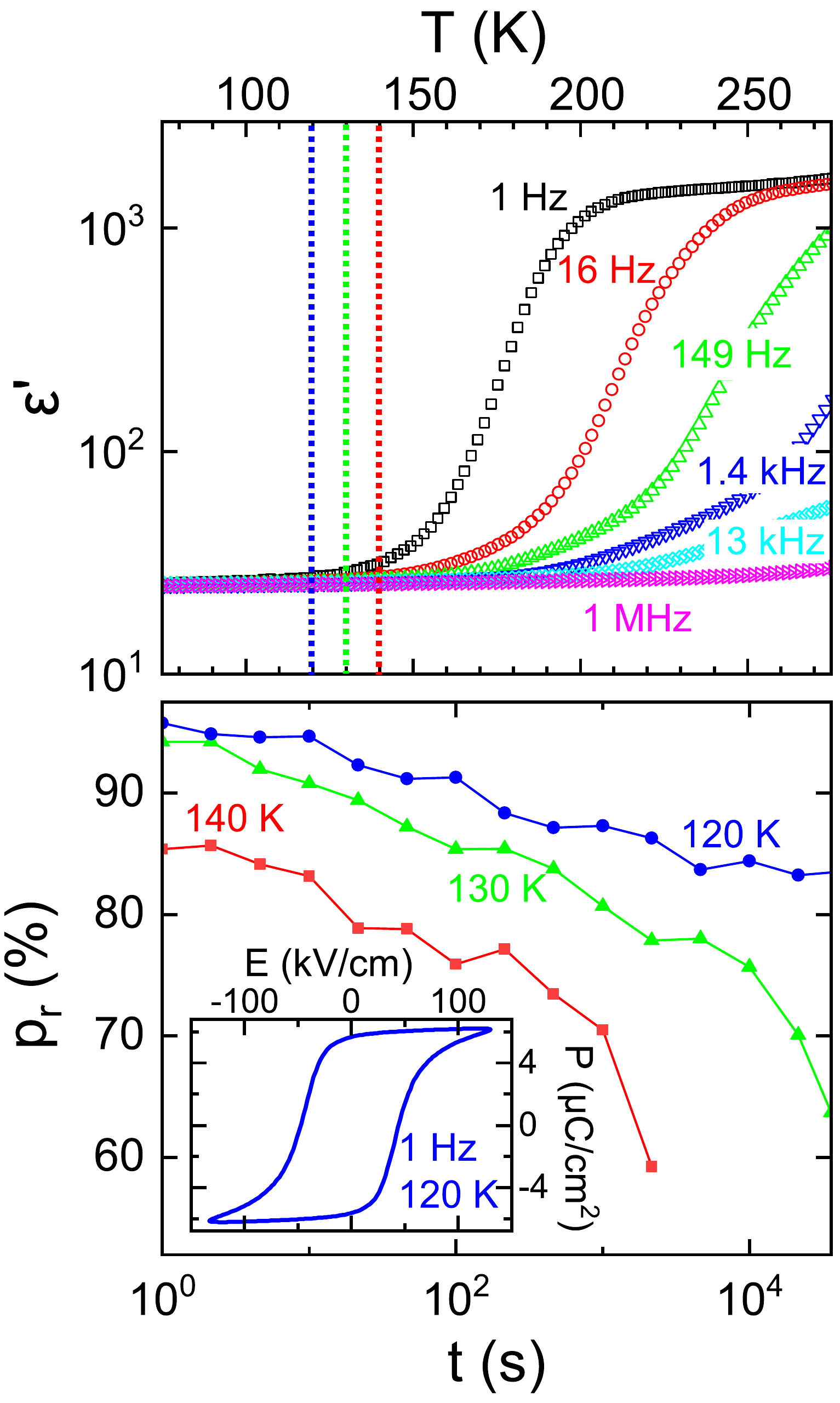}
		\caption{Spatially integrated bulk ferroelectric properties of  \YMO. a) Temperature-dependent dielectric constant  for selected frequencies measured by dielectric spectroscopy. The dashed lines denote the temperatures below which the intrinsic ferroelectric properties of the sample can be measured. b) Time-dependent  decay of the saturated polarization ($p_\mathrm{r}(t) = P_{\mathrm{meas}}(t)/P_{\mathrm{sat}}$, see text). At \SI{120}{\kelvin}, the polarization  is retained for several days, whereas at \SI{140}{\kelvin} $p_\mathrm{r}(t)$ relaxes towards equilibrium, \ie $p_\mathrm{r}=$\,50\%, within a few hours. Inset: ferroelectric hysteresis loop measured at \SI{120}{\kelvin} and \SI{1}{\hertz}.}
		\label{fig:dielectric}
	\end{figure}

	Complementary to the local measurements, we  performed bulk dielectric spectroscopy and ferroelectric hystersis loop measurements to characterize the retention of the  \YMO polarization and test for signatures of back-switching at the macro-scale.   Measurements  of the dielectric constant $\epsilon'$ shown in  Fig. \ref{fig:dielectric}\,a) revealed a step-like increase of $\epsilon'$ with temperature, indicating an intrinsic dielectric constant  masked by barrier layer capacitance effects\cite{Lunkenheimer2002,Lunkenheimer2009}.  Therefore, we chose our measurement temperature  such that we could probe the intrinsic ferroelectric polarization\cite{Lunkenheimer2009,Ruff2017,Ruff2018} (left of the dashed lines in Fig.~\ref{fig:dielectric}\,a)). Specifically, we performed all experiments at temperatures at or below 140~K. For confirmation,  we measured a ferroelectric hysteresis loop at \SI{120}{\kelvin} with an electrical poling field oscillating at \SI{1}{\hertz} (inset of Fig.~\ref{fig:dielectric}\,b). The shape of the loop and the saturation polarization are in perfect agreement with theory\cite{Fennie2005} and values of previous experiments \cite{Ruff2018,Han2013,Choi2010, Lilienblum2015}, confirming that only the true polarization was measured. 
	
	 To measure the retention behavior, first a pre-poling pulse with an applied electric field of \SI{120}{\kilo \volt \per \centi \meter} was used to saturate the sample polarization. After a delay time ranging from \SI{1}{\second} to \SI{3.6E5}{\second},  positive-up-negative-down (PUND) measurements with the first pulse in the same electrical-field direction as the pre-poling pulse and a peak electric field of \SI{120}{\kilo \volt \per \centi \meter} were performed to determine the  remaining fraction of the saturated polarization $p_\mathrm{r} (t) = P_{\mathrm{meas}}(t)/P_{\mathrm{sat}}$, where $P_{\mathrm{sat}}$ denotes the initial polarization created by the pre-poling pulse  and $P_{\mathrm{meas}}$ the measured polarization after the delay time $t$. Figure \ref{fig:dielectric}\,b) shows $p_\mathrm{r}$ as a function of the delay time $t$  measured at three different temperatures. The equilibrium state towards which the system relaxes corresponds to $p_\mathrm{r} = 50\%$, \ie equal amount of   up- and down polarized regions. At \SI{140}{\kelvin}, the polarization reverted quickly to this equilibrium state after poling, whereas the value of polarization surplus was retained for several days at \SI{120}{\kelvin}. These results are consistent with the reversal of polarization upon heating observed by the local switching experiments in Figs.~\ref{fig:square} and \ref{fig:bubble}.

\begin{figure}	
		\centering
		\includegraphics[width=\columnwidth]{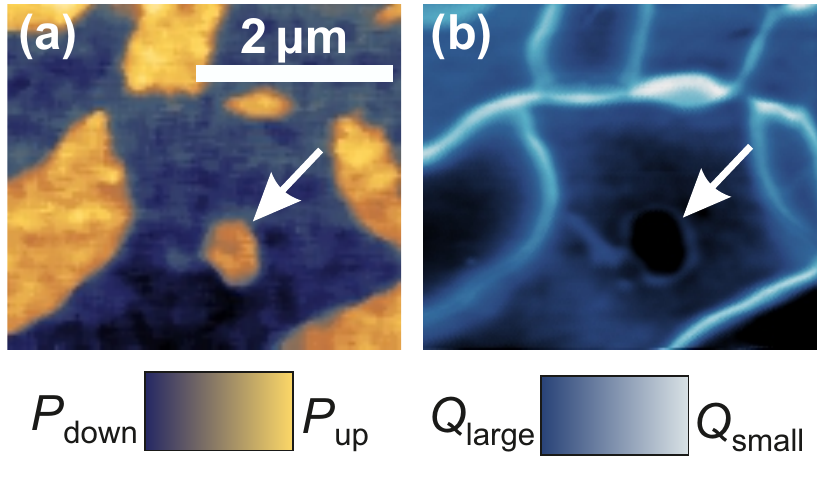}
		\caption{Electrostatic contrast at as-grown and electric-field-induced domain walls at \SI{120}{\kelvin}. a) PFM scan of the sample surface. An electric-field-induced bubble domain is highlighted by the white arrow. b)  EFM of the same area as in (a). Even though the PFM contrast is the same for both as-grown and electric-field-induced domains, the EFM contrast of the respective domain walls differs strongly between the two differently generated domains.}
		\label{fig:EFM}
	\end{figure}

	The domain walls  of as-grown and electric-field-induced polar domains also showed different behavior  when observed in EFM measurements. Because the overall conductivity is very low at \SI{120}{\kelvin}, the domain wall conductance can not directly be measured by conductive AFM. However, EFM allows to image the electrostatics of domain walls even under insulating conditions (see Ref. \onlinecite{Schoenherr2019} and Supplementary Information for details).  Fig.~\ref{fig:EFM}\,a) shows a PFM scan of the sample surface where a bubble domain was created by poling at \SI{120}{\kelvin} (arrow). Here, in contrast to the  measurements 
	in which the AFM tip was scanned over a defined area with an applied voltage,  the tip was stationary on the sample surface while applying the writing voltage. This resulted in the creation of a domain of about \SI{300}{\micro \meter} diameter as shown in Fig.~\ref{fig:EFM}\,a).  
	
	The as-grown and the electric-field-induced domains exhibited the same brightness in PFM. Figure~\ref{fig:EFM}\,b) shows an EFM image of the same area, measured at \SI{120}{\kelvin} after the sample had been heated to \SI{200}{\kelvin}. This temperature sequence creates an EFM contrast due to the pyroelectric effect associated with the temperature change, but preserves the written domain pattern because the temperature is not high enough for fast relaxation (see Supplementary Information for details). A pronounced EFM contrast was observed at as-grown domain walls, which is consistent with their enhanced conductivity attributed to the presence of oxygen interstitials\cite{Schaab2018}. At the  domain walls associated with the written domains, however, the EFM contrast was substantially weaker, suggesting lower electronic conductance and, hence, a lower density of oxygen defects compared to the as-grown walls.  
	
	\section{Discussion}
	
		\begin{figure}	
		\centering
		\includegraphics[width=\columnwidth]{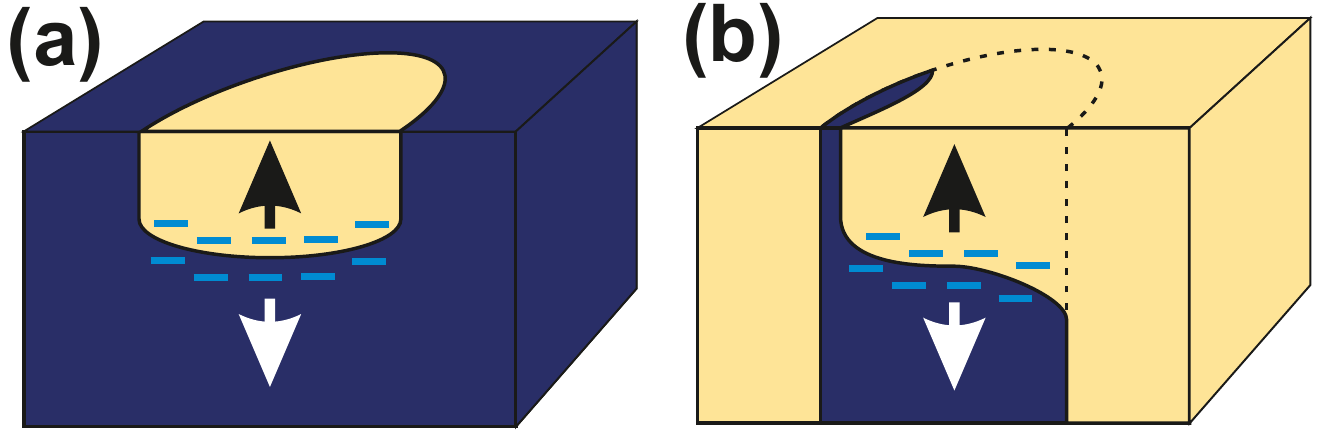}
		\caption{Schematic cross section of tip-electric-field-induced domain configurations and distribution of uncompensated charges ($-$). Arrows denote the polarization direction of the respective domains. a) Creation of a new domain at the surface, as described in Fig.~\ref{fig:square}. b) Deleting a bubble domain from the surface, as described in  Fig.~\ref{fig:bubble}. Vertical dimensions not to scale.}
		\label{fig:discussion}
	\end{figure}

	We now discuss why electric-field-induced domains tend to return to their as-grown, trimerization-controlled configuration upon heating. First, we note that even though the electric-field poling acts on the polarization, the trimerization has to follow this reorientation because of the rigid coupling between secondary and primary order parameter\cite{Kumagai2013}. Thus, we can exclude that the observed backswitching is due to an unswitched residue of the trimerized state. 
	
When the polarization at the sample surface is locally influenced by the AFM tip, it is affected only in a layer of a few hundreds of nanometers at the surface due to field-focusing below the AFM tip. Hence,  the  bulk polarization  below the field-induced square domain in Fig.~\ref{fig:square}  remained unswitched. At the newly created domain wall below, the polarizations  met tail-to-tail, resulting in uncompensated charges as is illustrated in Fig.~\ref{fig:discussion}\,a). At elevated temperatures, such a configuration would be readily screened by charge carriers, but at the cryogenic temperature of our experiment, this screening process becomes extremely slow.  Since the presence of uncompensated charges at the domain wall is energetically unfavorable,   the material  returned to its initial configuration when heated.
	
As-grown  domains, on the other hand,  exhibit no uncompensated charges and are therefore stable. An as-grown domain wall which had been erased by external electric fields was restored to its original shape by a temperature increase (Fig.~\ref{fig:bubble}). We conclude that defects, which show a propensity to accumulate at domain walls at high temperatures, but are immobilized at low temperatures\cite{Remsen2011}, could remain at their original location when a domain wall is displaced and serve as localized potential energy minima for the recovery of the domain structure. This hypothesis is corroborated by the remanent outline of the erased domain in Fig.~\ref{fig:bubble}\,b) and the difference in domain-wall contrast between as-grown and electric-field-induced domain walls in Fig.~\ref{fig:EFM}\,b), which can both be explained by a difference in defect density.  Note that a similar dissociation of domain walls and defects during switching was previously found in BiFeO$_3$ \cite{Stolichnov2014}.

The intriguing consequence of these conclusions is that the electrostatic forces in the improper ferroelectric \YMO are strong enough to  reverse not only the secondary, but also the primary order parameter, leading to the striking situation that the  allegedly weaker secondary order parameter controls the stronger one.
	
Note that in all  our local probe experiments, the topological protection of the domain structure by the primary order parameter did not play a role, because only bubbles, \ie domains within existing domains, were created and erased, whereas the topological domain vortex meeting points were not affected. Therefore, we observed a behavior resembling that of proper ferroelectrics. 

In our bulk switching experiments,  on the other hand, the topological constraints imposed by the primary order parameter affected the poling behavior. Specifically, electric-field poling cannot destroy the topological domain vortices and therefore the sample cannot be transferred into a single-domain state \cite{Choi2010, Jungk2010}. These unswitched remnants of the unfavored polarization direction then served as nuclei and memory in the relaxation of the polarization, a behavior not expected in proper ferroelectrics.

	\section{Summary and Conclusions}

We have shown that despite their origin in a non-polar, primary order parameter, the manipulation of polar domains in improper ferroelectric \YMO is guided by the same electrostatics as in proper ferroelectrics. In particular, the improper ferroelectric  domain configuration can be manipulated by electric fields, and its dynamics upon heating is driven by charge accumulation. On the other hand, bulk measurements indicate that the topological protection of the domain configuration due to the primary order parameter prevents the sample from reaching the single-domain configuration of a proper ferroelectric, with consequences for the nucleation, pinning and conductance of the remaining ferroelectric domain walls. We thus conclude that with regard to external fields and charges, improper ferroelectrics behave like a proper ferroelectric in many respects, but the existence of the primary order parameter leads to intriguing additional functionalities.

	\section{Acknowledgments}
	
	The authors thank M. C. Weber for valuable help in sample preparation and A. Bortis, D. M. Evans and Q. N. Meier for helpful discussions. This research	was supported by the EU European Research Council 	(Advanced Grant No. 694955—INSEETO) and the Swiss National Fund under grant numbers SNSF 20021\_178825, 20021\_149192 and 206021\_150635. L.K. acknowledges support from an ETH Career Seed Grant. D.M. thanks NTNU for support through the Onsager Fellowship Program and the Outstanding Academic Fellows Program.
	
	All authors discussed the results and contributed to the completion of the manuscript. L. K. and P. S. performed the low-temperature AFM experiments. S. K. and K. H. performed the dielectric spectroscopy measurements. E.P. grew the \YMO samples. S.K., T.L, M.T., D. M. and M.F. designed the experiment and supervised the study. 
	\bibliographystyle{apsrev4-1}
	\bibliography{../bibliography}

\end{document}


\preprint{APS/123-QED}

\title{Local control of improper ferroelectric domains in \YMO  \\ Supplementary Information}

\author{Lukas Kuerten}
\email{lukas.kuerten@mat.ethz.ch}
\affiliation{Departement of Materials Science, ETH Zurich,8093 Zurich, Switzerland}%
\author{Stephan Krohns}
\affiliation{Experimental Physics V, University of Augsburg, 86135 Augsburg, Germany}%
\author{Peggy Schoenherr}
\affiliation{Departement of Materials Science, ETH Zurich,8093 Zurich, Switzerland}%
\author{Katharina Holeczek}
\affiliation{Experimental Physics V, University of Augsburg, 86135 Augsburg, Germany}%
\author{Ekaterina Pomjakushina}%
\affiliation{Laboratory for Multiscale Materials Experiments, Paul Scherrer Institut, 5232 Villigen, Switzerland}
\author{Dennis Meier}%
\affiliation{Department of Materials Science and Engineering, Norwegian University of Science and Technology, 7043 Trondheim, Norway}
\author{Thomas Lottermoser}\author{Morgan Trassin}\author{Manfred Fiebig}
\affiliation{Departement of Materials Science, ETH Zurich,8093 Zurich, Switzerland}%

%
%

\date{\today}

\maketitle


\section{Imaging of surface charges with electrostatic force microscopy}

\begin{figure}
	\centering
	\includegraphics[width=0.4\textwidth]{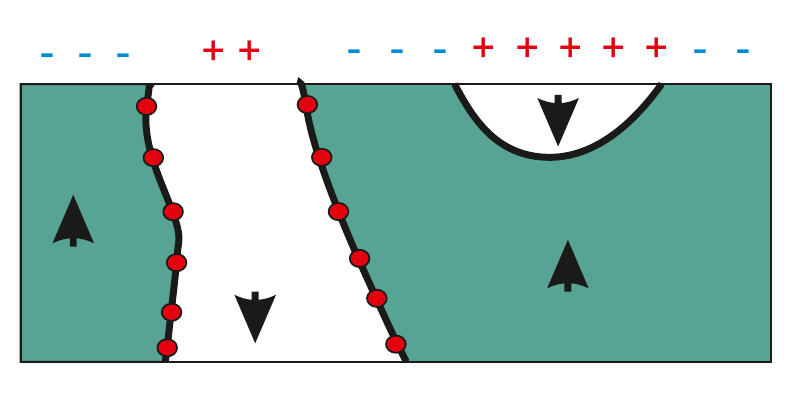}
	\caption{Schematic illustration: as-grown domain walls are decorated with defects (red dots), which facilitate conductivity and can drain  surface charges, resulting in uncharged regions on the sample surface. Electric-field-induced domain walls do not contain defects, therefore the surface charge is not drained. Because the absolute value of the charge does not change at electric-field-induced domain walls, they do not appear in the EFM image when the tip is grounded.}
	\label{figB}
\end{figure}

Ferroelectric domains can be imaged by electrostatic force microscopy (EFM) because when the polarization of the sample changes, non-equlibrium conditions can result in over- or underscreened surfaces with a net surface charge \cite{Kalinin2001, Schoenherr2019}. At ambient conditions, polarization and charge reach equilibrium quickly because there are efficient mechanisms of charge transport. At low temperatures and in inert atmosphere, however, relaxation mechanisms are slow and non-equlibrium conditions on the sample surface can persist for extended periods of time. In particular, when the polar material undergoes a temperature change, uncompensated charges at the surface can be detected by EFM \cite{Schoenherr2020}. In semiconducting materials like \YMO, intrinsic conductivity is relevant at room temperature but at low temperatures only defect conductivity plays a significant role \cite{Sze2007}. Therefore, surface charging is significantly different between regions of high and low defect density. In the present case, the as-grown domain walls exhibit a high density of defects, whereas domains and electric-field-induced domain walls have a low density. This leads to a difference in surface charge and allows to distinguish these features in the EFM contrast, as illustrated in figure \ref{figB}. In the EFM mode used here, the conducting tip is grounded and therefore does not distinguish between positive or negative charges, therefore the EFM signal depends solely on the absolute value of the surface charge.

\begin{figure}
	\centering
	\includegraphics[width=0.4\textwidth]{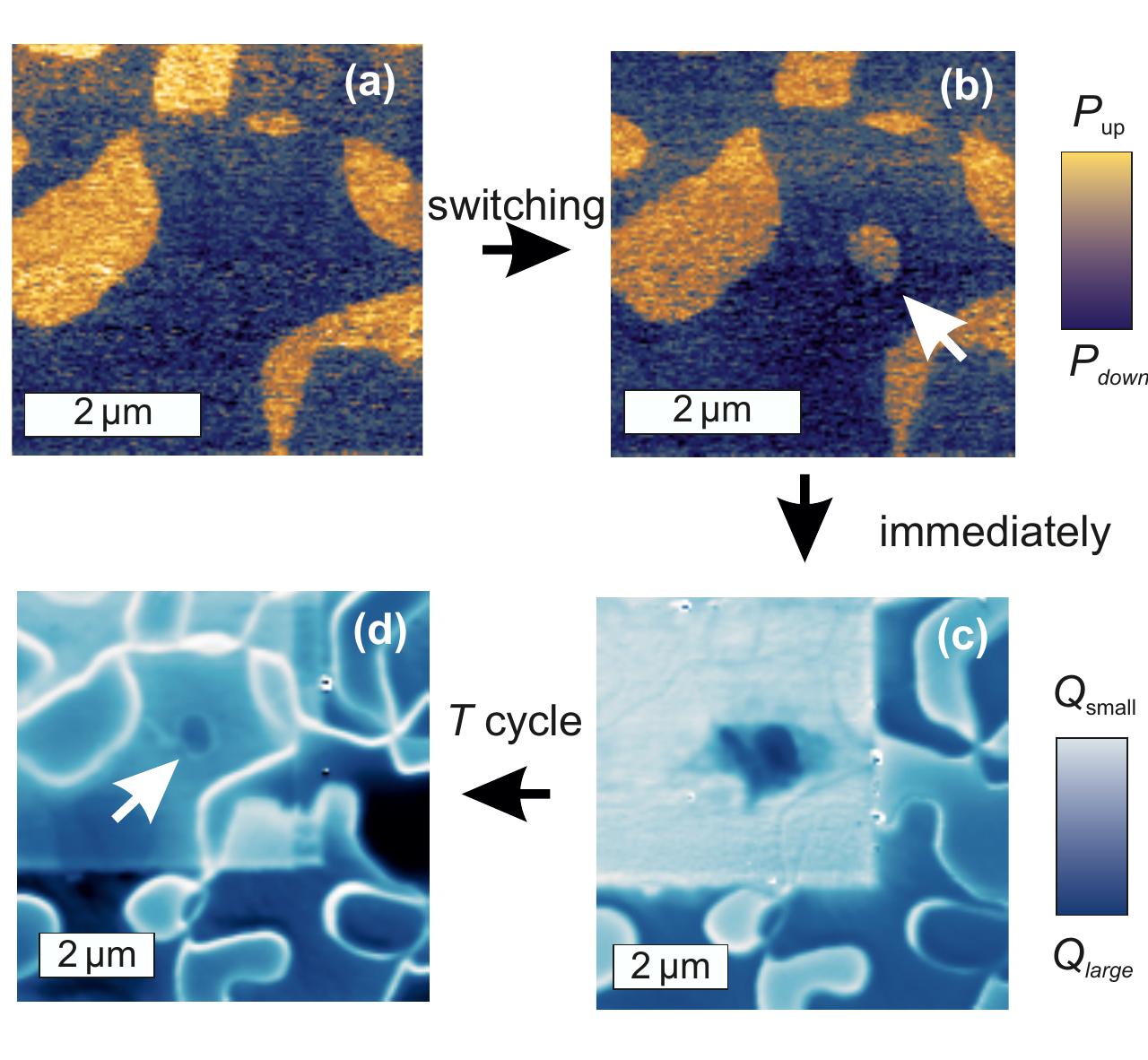}
	\caption{Illustration of the prodecure used to obtain the PFM and EFM images shown in figure 3. All images taken at \SI{120}{\kelvin}.  Clockwise from the top left: a) PFM image of the pristine domain structure of the sample. b) PFM image after  creating a bubble of upwards polarization by field-induced switching, highlighted by the white arrow. c) EFM image of a larger area, measured directly after the PFM scan. In the region of the PFM scan, the tip scanning in contact has removed charges from the sample surface, thereby destroying EFM contrast. The surrounding area is unaffected and still displays domain wall contrast. d) The sample was heated to \SI{200}{\kelvin} and cooled back to \SI{120}{\kelvin}. This temperature cycle does not affect the electric-field-induced domain, but it restores the charges on the sample surface, resulting in renewed EFM contrast also in the PFM scan area.  Note the difference in lateral scale between the PFM and EFM images.}
	\label{figA}
\end{figure}

When a PFM scan or lithography is performed on the sample surface, scanning the AFM tip over the surface in contact  erases the screening charges and destroys the EFM contrast in that section of the sample surface. Therefore, in order to image an area with EFM after a contact-mode scan was performed in the same area, a temperature cycle has to be performed between contact-mode and EFM measurement. In order to recreate EFM contrast, the sample is heated to \SI{200}{\kelvin} and cooled back to \SI{120}{\kelvin}. This increase in temperature does not trigger relaxation of the imprinted domain, but it is sufficient to induce surface charges generating an EFM contrast This temperature sequence is illustrated in figure \ref{figA}.

\bibliographystyle{apsrev4-1}
\bibliography{bibliography}